\documentclass[12pt, a4paper]{article}
\title{My first LaTeX document}

\usepackage{authblk}
\title{Simultaneous estimation of electrical conductivity and permittivity in quantitative thermoacoustic tomography}
\author[1]{Teemu Sahlström}
\author[1]{Timo Lähivaara}
\author[1]{Tanja Tarvainen}
\affil[1]{Department of Technical Physics, University of Eastern Finland, Kuopio, Finland}
\date{}           
\usepackage{amsmath}
\usepackage{amsfonts}
\usepackage{mathtools}
\usepackage{amssymb}

\begin{document}
\maketitle
\begin{abstract}
In this work, the inverse problem of quantitative thermoacoustic tomography is studied. In quantitative thermoacoustic tomography, dielectric parameters of an imaged target are estimated from an absorbed energy density induced by an externally introduced electromagnetic excitation. In this work, simultaneous estimation of electrical conductivity and permittivity is considered. We approach this problem in the framework of Bayesian inverse problems. The dielectric parameters are estimated by computing \emph{maximum a posteriori} estimates, and the reliability of the estimates is studied using the Laplace's approximation. The forward model to describe electromagnetic wave propagation is based on a vectorial Maxwell's equation, that is numerically approximated using a finite element method with edge elements. The proposed methodology is evaluated using numerical simulations utilising one or two electromagnetic excitations at multiple excitation frequencies. The results show that the electrical conductivity and permittivity can be simultaneously estimated in quantitative thermoacoustic tomography. However, the problem can suffer from non-uniqueness, which could be overcome using multiple electromagnetic excitations.
\end{abstract}
	
	\vspace{2pc}
	
\section{Introduction}

In quantitative thermoacoustic tomography (QTAT), the aim is to estimate the dielectric parameters of an imaged target from boundary measurement of ultrasound waves generated by the thermoacoustic effect \cite{li2008,li2009,bal2011,akhoyari2017}. In QTAT experiments, short pulses of  micro- or radio waves (typically from 300 MHz to 3 GHz) are directed to the imaged target \cite{cui2017review}. As the electromagnetic wave propagates through the imaged target, its energy is absorbed leading to temperature rise and localised increases in pressure. This pressure increase relaxes as broadband ultrasound waves, that are measured on the boundary of the target.

Similar to quantitative photoacoustic tomography (QPAT), QTAT is an imaging modality based on coupled physics. In QPAT, the ultrasound signals are generated by absorption of an externally induced short pulse of visible or near-infrared light \cite{li2009}. The optical parameters, such as absorption and scattering, are then estimated from the measured ultrasound signals \cite{li2009,cox2012,tarvainen2024quantitative}. Due to the relatively high absorption of light in biological tissues, imaging depth of QPAT is limited. In contrast, QTAT can achieve greater imaging depths because of the relatively high penetration depth of micro- and radio waves \cite{li2009}. The main benefit of QPAT and QTAT, and coupled physics imaging in general, is the capability of combining the contrast and resolution of different physical phenomena. In QTAT, the unique contrast of electromagnetic waves is combined with the high resolution of ultrasound.

A complete forward model and a formal solution of the inverse problem of QTAT, including the electromagnetic and acoustic models and taking into account slowly time-varying electromagnetic excitation, was presented in \cite{akhoyari2017}. Furthermore, an algorithm for reconstructing the conductivity directly from ultrasound waves was presented in \cite{al2020direct}, and an approach for estimating the electrical and thermal conductivities was studied in \cite{cristofol2023carleman}. However, numerical implementations of the complete problem were not presented. 

Generally, forward modelling and inverse problems of QPAT and QTAT are approached in two stages: acoustic and optical or acoustic and electrical, respectively. However, although this approach is generally employed in the inverse problem of QTAT, it should be noted that authors in \cite{akhoyari2017} showed that, due to the longer pulse of induced microwave and resulting slower absorption and thermoacoustic effect, the two parts cannot  always be decoupled. In the two-stage approach for the inverse problem of QTAT, the acoustic inverse problem amounts to estimating the absorbed energy density from the boundary measurements of ultrasound waves. In the electrical inverse problem, the dielectric properties of the imaged target are estimated from the absorbed energy density. Modelling and inverse problem of the acoustic part of QTAT, typically referred as TAT, have been studied widely \cite{wang2011bookchapter,kuchment2011bookchapter,poudel2019survey}, and several experimental systems have been developed \cite{kruger1999thermoacoustic,wang1999microwave,fallon2009rf,razansky2010near,ogunlade2012quantitative,cui2017review}. In this work, we consider the electrical inverse problem of estimating the dielectric parameters from the absorbed energy density, and refer to it as the inverse problem of QTAT. We further assume that the absorbed energy density, i.e. the solution to the acoustic inverse problem, is known. 

There are only few studies that have considered the estimation of the dielectric parameters in QTAT, and even fewer where the problem has been studied with numerical simulations. Estimating the electrical conductivity, when the permittivity was assumed to be known, was studied in \cite{bal2011,ammari2013}. In \cite{bal2011}, it was shown, that the conductivity can be uniquely and stably estimated from one absorbed energy density measurement when the conductivity is sufficiently small compared to the frequency of the electric field and the permittivity is a constant. The approach was studied with numerical simulations by estimating the conductivity using both the scalar Helmholtz equation and the vectorial Maxwell's equation. In \cite{ammari2013}, an analytical reconstruction formula for estimating the conductivity using the Helmholtz equation was presented. When considering estimation of both conductivity and permittivity, it has been shown that they can be uniquely and stably reconstructed using a set of well-chosen electromagnetic excitations \cite{bal2014}. Furthermore, in \cite{ren2023recovering} it was shown that, under appropriate conditions, the dielectric parameters and the Gr{\"u}neisen parameter, that is used to describe thermoacoustic efficiency, can be estimated from an initial pressure based on a system of semilinear Helmholtz equations. The methodology was evaluated using numerical simulations. 

In this work, we propose an approach for simultaneous estimation of the conductivity and permittivity in QTAT. The forward problem is formulated using a vectorial Maxwell's equation with an impedance boundary condition to model incident electric fields. Its solution is approximated numerically with a finite element method (FEM) using edge elements. In the inverse problem, we solve the \emph{maximum a posteriori} (MAP) estimate using the the Gauss-Newton method. Furthermore, the reliability of the MAP estimates is evaluated utilising the Laplace's approximation. 

The remainder of the paper is organised as follows. The forward model and numerical simulation of electric fields using the FEM is presented in Section \ref{sec:forward_problem}. The inverse problem and its numerical solution are presented in Section \ref{sec:inverse_problem}. The simulations are shown in Section \ref{sec:simulations}. Finally, the results are discussed and conclusions are given in Sections \ref{sec:discussion} and \ref{sec:conclusions}, respectively.

\section{Forward problem}
\label{sec:forward_problem}

Let us consider a domain $\Omega \in \mathbb{R}^d$ ($d = 2,3$) with a boundary $\partial \Omega$. In the forward problem of QTAT, the absorbed energy density $H(r,t)$, where $r\in\Omega$ is a spatial position and $t$ denotes time, is solved when dielectric material parameters and an electromagnetic source are given. In this work, we assume linear, isotropic, non-magnetic and inhomogeneous materials with a conductivity $\sigma(r) \geq 0$ and a relative permittivity $\epsilon_r(r) \geq 1$. We consider time-harmonic electromagnetic fields of the form $\hat{E}(r,t) = \mathrm{Re}(E(r)e^{\mathrm{i}\omega t})$ with $E(r) \in \mathbb{C}^d$, where $\mathrm{i}$ is the imaginary unit. The complex amplitude $E(r)$ can then be described by the time-harmonic double-curl Maxwell's equation \cite{book_monk,book_harrington}
\begin{eqnarray}
	\nabla \times \nabla \times E(r) - \gamma^2(r) E(r) = 0, \quad r \in \Omega,
	\label{eq:double_curl}
\end{eqnarray}
where $\gamma(r) = \sqrt{\mu_0 ( \omega^2\epsilon_0\epsilon_r(r) - \mathrm{i} \omega \sigma(r) )}$, $\mu_0$ and $\epsilon_0$ are the permeability and permittivity  of free space, $\omega = 2\pi \tilde{f}$ is the angular frequency, and $\tilde{f}$ is the frequency. Furthermore, we assume the exterior of $\Omega$ to have a constant conductivity $\sigma_{\rm ext}$ and a constant relative permittivity $\epsilon_{r,{\rm ext}}$. 

To introduce incident electromagnetic waves to $\Omega$, an impedance boundary condition is set on the boundary $\partial \Omega$ leading to  a system of equations \cite{book_monk,thesis_bonazolli, thesis_watson}
\begin{equation}
	\begin{dcases}
		\nabla \times \nabla \times E(r) - \gamma^2(r) E(r) = 0, \quad r \in \Omega,\\
		(\nabla \times E(r)) \times \hat{n} + \mathrm{i}\kappa \hat{n} \times (E(r) \times \hat{n}) = g(r),  \quad r \in \partial \Omega \label{eq:double_curl_impedance_boundary} 
		\end{dcases}
\end{equation}
where $\kappa = \sqrt{\mu_0 ( \omega^2\epsilon_0\epsilon_{r,{\rm ext}} - \mathrm{i} \omega \sigma_{\rm ext} )}$, $\hat{n}$ is an outward unit normal, and $g(r)$ is dependent on the incident electric field $E_{{\rm inc}}$.

As electromagnetic waves propagate through the medium, they are absorbed, resulting in an absorbed energy density $H(r,t)$. The absorbed energy density can be described using the specific absorption rate (SAR) as \cite{li2008,li2009}
\begin{eqnarray}
	H(r,t) = \rho(r) \mathrm{SAR}(r,t),
	\label{eq:H_time}
\end{eqnarray}
where $\rho(r)$ is the density of the material, and
\begin{eqnarray}
	\mathrm{SAR}(r,t) = \frac{\sigma (r) \vert \hat{E}(r,t) \vert^2 }{\rho (r)}.
	\label{eq:SAR}
\end{eqnarray}

If the electromagnetic pulse is sufficiently short compared to the time scale of thermal diffusion, the absorbed energy density can be separated in spatial and temporal components as $H(r,t) = H(r)\tilde{I}(t)$, where $\tilde{I}(t)$ is a temporal envelope of the spatial component $H(r)$. Furthermore, when the absorption of electromagnetic energy is assumed to be instantaneous, the temporal term can be modelled as a Dirac delta function and the absorbed energy density is given by \cite{li2008,li2009,bal2011}
\begin{eqnarray}
	H(r) = \sigma(r) \vert E(r) \vert^2.
	\label{eq:H_instant}
\end{eqnarray}

\subsection{Finite element approximation}
\label{subsec:finite_element_approximation}

In this work, solution $E(r)$ of \eqref{eq:double_curl_impedance_boundary} is approximated using the finite element method (FEM). We seek the solution in the function space $\Psi = \{\psi \in H^{\mathrm{curl}}(\Omega)\}$ of square integrable functions with square integrable curls. By multiplying the first equation in equation \eqref{eq:double_curl_impedance_boundary} with a test function $\psi(r) \in \Psi$, integrating over the domain $\Omega$, and utilising the boundary condition, the variational problem is to find $E(r) \in \Psi$ such that \cite{book_monk,book_li}
\begin{align}
	\int_\Omega (\nabla \times & E(r)) \cdot (\nabla \times {\psi(r)}) \, d\mathrm{r} - \int_\Omega \mu_0(\omega^2\epsilon_0\epsilon_r(r) - \mathrm{i}\omega\sigma(r)) E(r) \cdot {\psi(r)} \, d\mathrm{r} \nonumber   \\
	& + \int_{\partial\Omega} \mathrm{i} \kappa(E(r)\times \hat{n}) \cdot ({\psi(r)} \times \hat{n}) \, d\mathrm{r} = \int_{\partial\Omega} g(r) \cdot {\psi(r)} \, d\mathrm{r}.
	\label{eq:variational_form}
\end{align}

Following the approach described in \cite{thesis_bonazolli, thesis_watson}, we represent the solution $E(r)$ of the variational form \eqref{eq:variational_form} using linear N\'ed\'elec edge elements in a triangular mesh. Further, we use the same basis for the test functions. The edge elements are especially suitable for electromagnetic problems due to their ability to model continuity properties of electric fields on material boundaries, and the ability to avoid non-physical numerical solutions \cite{thesis_bonazolli,book_monk,book_li}. 

For linear N\'ed\'elec elements, the degrees of freedom are associated with edges of the mesh elements, and the electric field can be written as a linear combination of vector-valued basis functions $\psi_i(r)$ as
\begin{eqnarray}
	E(r) \approx E_h(r) = \sum_{i=1}^{I} \varepsilon_i \psi_i(r),
	\label{eq:E_FEM_discretisation}
\end{eqnarray}
where $I$ denotes the number of element edges in the domain $\Omega$. We further discretise the dielectric parameters as piece-wise (element-wise) constant
\begin{eqnarray}
	\sigma(r) \approx \sigma_h(r) = \sum_{j = 1}^{J} \sigma_j \chi_j(r) \quad \mathrm{and} \quad \epsilon_r(r) \approx \epsilon_{r,h}(r) = \sum_{j = 1}^{J} \epsilon_{r,j} \chi_j(r),
\end{eqnarray}
where $\chi_j$ is a characteristic function of the $j$:th element and $J$ denotes the number of elements. 
The electric field coefficients on the element edges $\varepsilon = (\varepsilon_1, \ldots , \varepsilon_I^{\rm T}) \in \mathbb{C}^I$ can be solved from a linear system of equations of the form
\begin{eqnarray}
	K\varepsilon = b.
	\label{eq:FEM_system}
\end{eqnarray}
where $K = S + M + R$. The system matrices $S$, $M$, and $R$, and the vector $b$ are given by
\begin{align}
	S(m,n)   &= \int_\Omega  (\nabla \times \psi_m(r)) \cdot (\nabla \times \psi_n(r)) \, \mathrm{d}r  \label{eq:FEM_system_S}\\
	M(m,n)   &= \sum_{j = 1}^{J} -\mu_0(\omega^2\epsilon_0 \epsilon_{r,j} - \mathrm{i}\omega\sigma_j) \int_{\Omega_j} \psi_m(r) \cdot \psi_n(r) \, \mathrm{d}r \label{eq:FEM_system_M} \\
	R(m,n)   &= \int_{\partial\Omega} \mathrm{i}\kappa(\psi_m(r) \times \hat{n}) \cdot (\psi_n(r) \times \hat{n}) \, \mathrm{d}r \label{eq:FEM_system_R} \\
	b(m)     &= \int_{\partial\Omega} g(r) \cdot \psi_m(r) \, \mathrm{d}r, \label{eq:FEM_system_b}
\end{align}
where $m,n = 1, ..., I$ and $\Omega_j$ denotes integration over the $j$:th element \cite{thesis_bonazolli,anjam2015}. 

For a discretised representation of the data, the absorbed energy density is presented in a piece-wise constant basis 
\begin{eqnarray}
	H = \mathrm{diag}\{ \sigma\} \vert E \vert ^2 = \mathrm{diag}\{ \sigma\} \vert \mathcal{M}\varepsilon \vert ^2 .
	\label{eq:H_FEM}
\end{eqnarray}
where $H = (H_1, \dots, H_J)^\mathrm{T}$,  $\mathrm{diag}\{ \sigma \}$ is a diagonal matrix of conductivity coefficients $\sigma=(\sigma_1,\ldots,\sigma_J)^{\rm T}$,  $\mathcal{M}$  is a discrete measurement operator that maps the electric field coefficients from element edges to piece-wise constant presentation $E = (E_1, \dots, E_J)^\mathrm{T}$.

\section{Inverse problem}
\label{sec:inverse_problem}

In the inverse problem of QTAT, the conductivity $\sigma$ and relative permittivity $\epsilon_r$ of the medium are estimated from the absorbed energy density $y$. Let us denote the absorbed energy density data by a vector $y = (H^1, \ldots, H^L)^\mathrm{T} \in \mathbb{R}^{L}$ where $L = PJ$ is the number of data points and $P$ is the number of electromagnetic excitations. The observation model of QTAT with an additive noise model is of the form
\begin{equation}
	y = f(x) + e,
	\label{eq:forward_model}
\end{equation}
where  $f: \mathbb{R}^{2J} \rightarrow \mathbb{R}^L$ is a discretised forward operator mapping the dielectric parameters $x = (\sigma_1, \ldots, \sigma_J, \epsilon_{r,1}, \ldots ,\epsilon_{r,J})^\mathrm{T}  \in \mathbb{R}^{2J}$ to the data, and $e \in \mathbb{R}^{L}$ denotes the noise.

In this work, we approach the inverse problem of QTAT in the framework of Bayesian inverse problems \cite{book_calvetti, book_kaipio}. All parameters are modelled as random variables, and the solution of the inverse problem, that is the posterior distribution, is given by the Bayes' formula 
\begin{equation}
	\pi(x \vert y) \propto {\pi(y \vert x) \pi(x)},
\end{equation}
where $\pi(y\vert x)$ is the likelihood distribution and $\pi(x)$ is the prior distribution.

We model the unknown parameters $x$ and the noise $e$ as mutually independent and Gaussian distributed
\begin{eqnarray}
	x \sim \mathcal{N}(\eta_x, \Gamma_x), \quad e \sim \mathcal{N}(\eta_e, \Gamma_e), \nonumber
\end{eqnarray}
where $\eta_x$ and $\Gamma_x$ are the mean and covariance of the prior distribution and $\eta_e$ and $\Gamma_e$ are the mean and covariance of the noise. This leads to a posterior distribution of the form
\begin{eqnarray}
	\pi(x \vert y) \propto \exp\left\{ -\frac{1}{2} \Vert L_e(y-f(x)-\eta_e) \Vert^2 -\frac{1}{2} \Vert L_x(x- \eta_x) \Vert^2  \right\},
	\label{eq:posterior_distribution}
\end{eqnarray}
where $L_e$ and $L_x$ are the square roots of the inverse covariance matrices of the noise and prior, such as the Cholesky decomposition, $L_e^\mathrm{T}L_e = \Gamma_e^{-1}$ and $L_x^\mathrm{T}L_e = \Gamma_x^{-1}$, respectively.

Calculating the entire posterior distribution can be computationally prohibitively expensive in large dimensional tomography problems, and thus point estimates are often considered. In this work, we compute the \emph{maximum a posteriori} (MAP) estimate \cite{book_calvetti, book_kaipio}
\begin{eqnarray}
	x_{\mathrm{MAP}} = \arg \, \min _{x}  \left\{ \frac{1}{2} \Vert L_e(y-f(x)-\eta_e) \Vert^2 + \frac{1}{2} \Vert  L_x(x- \eta_x) \Vert^2 \right\}.
	\label{eq:map_optimisation}
\end{eqnarray}
This minimisation problem can be solved using methods of computational optimisation \cite{nocedal_book}. In this work, we use the Gauss-Newton method where the solution is computed iteratively as
\begin{equation}
	x_{l+1} = x_l + \alpha_l d_l,
	\label{eq:minimisation_iteration}
\end{equation}
where $\alpha_l$ is the step length and the search direction $d_l$ is
\begin{align}
	d_l = &\left(  J_{f(x_l)}^\mathrm{T} \Gamma_e^{-1} J_{f(x_l)} + \Gamma_x^{-1}  \right)^{-1} \nonumber\\
	&\left( J_{f(x_l)}^\mathrm{T} \Gamma_e^{-1} (y-f(x_l) - \eta_e) - \Gamma_x^{-1}(x - \eta_x) \right),
	\label{eq:GN_search_direction}
\end{align}
where $J_{f(x_l)}$ is the Jacobian matrix of the forward operator $f(x_l)$ at $x_l$ on the $l$:th iteration. 

We further evaluate the reliability of the MAP estimates by forming credibility intervals utilising the Laplace's approximation of the posterior distribution. For this, the forward operator is approximated using a Taylor series at $x_{\mathrm{MAP}}$ \cite{pulkkinen2016}
\begin{eqnarray}
	f(x) \approx f(x_{\rm MAP}) + J_{f(x_{\rm MAP})}(x - x_{\rm MAP}),
\end{eqnarray}
where $J_{f(x_{\rm MAP})}$ is the Jacobian matrix at $x_{\rm MAP}$. Covariance of the posterior distribution \eqref{eq:posterior_distribution} can then be approximated as \cite{book_tarantola,pulkkinen2016}
\begin{eqnarray}
	\Gamma_{x\vert y} = \left(J_{f(x_{\rm MAP})}^\mathrm{T} \Gamma_e^{-1} J_{f(x_{\rm MAP})} + \Gamma_x^{-1}\right)^{-1}.
	\label{eq:covariance}
\end{eqnarray}
The reliability of the MAP estimates can be studied by approximating credibility intervals. For example, the $99.7\, \% $  credibility interval for the posterior distribution can be approximated as a $\pm$3 standard deviation interval
\begin{eqnarray}
	[x_{\rm MAP} - 3\tilde{\sigma}_{x \vert y, j} \, , \, 	x_{\rm MAP} + 3\tilde{\sigma}_{x \vert y, j}],
	\label{eq:credibility_interval}
\end{eqnarray}
where $\tilde{\sigma}_j = \sqrt{\Gamma_{x \vert y,jj}}$ is the standard deviation of $x_j$.

\subsection{Jacobian matrix}
\label{subsec:jacobian_matrix}

The Jacobian matrix $J_f$ containing the partial derivatives of the forward operator with respect to the conductivity $\sigma_j$ and relative permittivity $\epsilon_{r,j}$ of a discretisation element $j=1,\dots,J$ for each electromagnetic excitation $p=1,\dots,P$, is of the form
\begin{eqnarray}
	J_f = [J_{\sigma}, J_{\epsilon_r}] = 
	\begin{pmatrix} \frac{\partial f^1}{\partial \sigma_1} & \dots & \frac{\partial f^1}{\partial \sigma_J} & \frac{\partial f^1}{\partial \epsilon_{r,1}}  & \dots, &  \frac{\partial f^1}{\partial \epsilon_{r,J}} \cr \vdots  &  & \vdots  & \vdots  & & \vdots \cr   \frac{\partial f^P}{\partial \sigma_1} & \dots & \frac{\partial f^P}{\partial \sigma_J} & \frac{\partial f^P}{\partial \epsilon_{r,1}}  & \dots, &  \frac{\partial f^P}{\partial \epsilon_{r,J}} \end{pmatrix}.
	\label{eq:Jacobian_matrix}
\end{eqnarray}
The derivatives of the forward operator $f^{p}$ with respect to conductivity $\sigma_j$ and relative permittivity $\epsilon_{r,j}$ are obtained as a column-wise vectorisation of
\begin{align}
	\frac{\partial f^p}{\partial \sigma_j} &=  \chi_j\vert E \vert^2   +  \sigma \odot \frac{\partial \vert E \vert ^2}{\partial \sigma_j} 	\label{eq:partial_f_sigma} \\ 
	\frac{\partial f^p}{\partial \epsilon_{r,j}} &= \sigma \odot \frac{\partial \vert E \vert ^2}{\partial \epsilon_{r,j}},
	\label{eq:partial_f_epsilon}
\end{align}
where $\odot$ is the element-wise Hadamard product. Further, for each electromagnetic excitation, the partial derivatives of the electric field are given by
\begin{align}
	\frac{\partial \vert E \vert ^2}{\partial \sigma_j} &= 2 \mathrm{Re}\left(\frac{\partial E}{\partial \sigma_j} \odot \overline{E}\right) = 2 \mathrm{Re}\left(\mathcal{M}\frac{\partial \varepsilon}{\partial \sigma_j} \odot \overline{\mathcal{M}\varepsilon}\right) \\
	\frac{\partial \vert E \vert ^2}{\partial \epsilon_{r,j}} &= 2 \mathrm{Re}\left(\frac{\partial E}{\partial \epsilon_{r,j}} \odot \overline{E}\right) = 2 \mathrm{Re}\left(\mathcal{M}\frac{\partial \varepsilon}{\partial \epsilon_{r,j}} \odot \overline{\mathcal{M}\varepsilon}\right),
\end{align}
where $\mathrm{Re}(\cdot)$ denotes the real part, $\overline{E}$ is the complex conjugate of $E$, and $\overline{\mathcal{M}\varepsilon}$ is the complex conjugate of $\mathcal{M}\varepsilon$. Furthermore, by differentiating the FE-system \eqref{eq:FEM_system} with respect to $\sigma_j$ and $\epsilon_{r,j}$, the partial derivatives of the electric field coefficients $\varepsilon$ are obtained from
\begin{align}
	&\frac{\partial \varepsilon}{\partial \sigma_j}    = -K^{-1} \frac{\partial M}{\partial \sigma_j} e = K^{-1} \mathrm{i} \mu_0\omega M_j \varepsilon \label{eq:partial_e_sigma} \\
	&\frac{\partial \varepsilon}{\partial \epsilon_{r,j}}  = -K^{-1} \frac{\partial M}{\partial \epsilon_{r,j}}  e = -K^{-1} \mu_0\epsilon_0\omega^2 M_j \varepsilon, 
	\label{eq:partial_e_epsilon}
\end{align}
where $M_j$ is the system matrix $M$ of the $j$:th element, equation \eqref{eq:FEM_system_M}.

\section{Simulations}
\label{sec:simulations}

\begin{figure}[tbp!]
	\centering
	\includegraphics[width=0.6\textwidth]{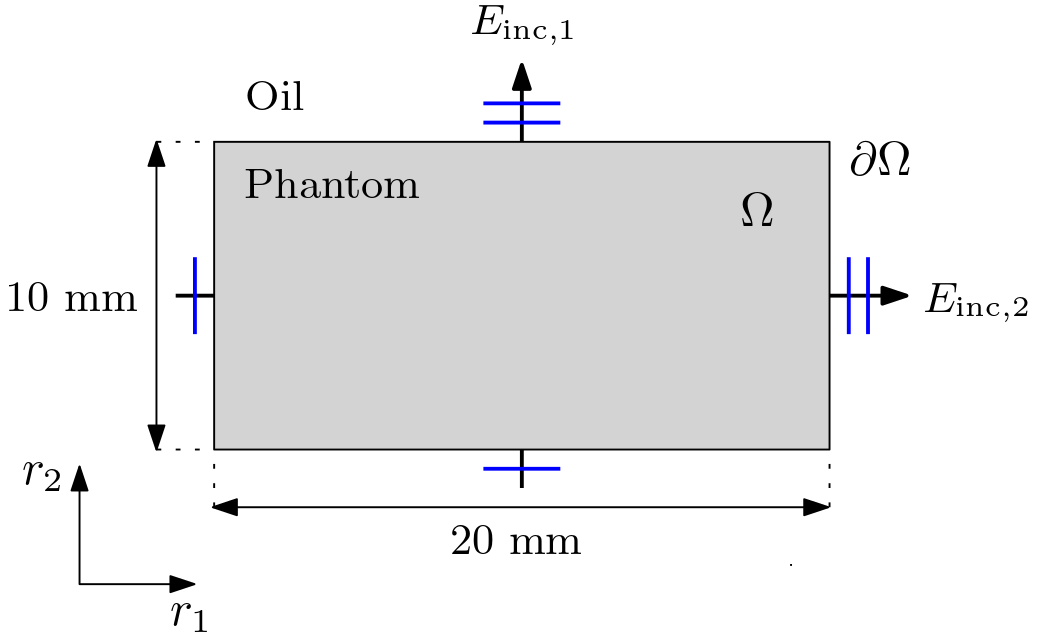}
	\caption{Illustration of the simulation setup. The simulation domain $\Omega$ is a $20 \, {\mathrm{mm}} \times 10 \, {\mathrm{mm}}$  rectangle. Exterior of the phantom is modelled as castor oil. The incident electric fields are plane waves $E_{{\mathrm{inc}},1}$ and $E_{{\mathrm{inc}},2}$ are travelling to $r_2$ and $r_1$ directions, respectively.   }
	\label{fig:simulation_illustration}
\end{figure}

The proposed approach for simultaneous estimation of electrical conductivity and relative permittivity was evaluated with numerical simulations in a $20 \, {\mathrm{mm}} \times 10 \, {\mathrm{mm}}$ rectangular domain. The exterior of the domain was modelled as castor oil with conductivity $\sigma_{\mathrm{ext}} = 1 \cdot 10^{-12} \, {\mathrm{Sm}}^{-1}$ and relative permittivity $\epsilon_{r,{\mathrm{ext}}} = 4$ (dimensionless). The electromagnetic excitation was simulated using the impedance boundary condition in equation \eqref{eq:double_curl_impedance_boundary}. For this, we considered two linearly polarised incident plane waves 
\begin{eqnarray}
	E_{{\mathrm{inc}},1}(r) =  \begin{pmatrix}  \exp\{-\mathrm{i} \kappa r_2\} \cr 0 \end{pmatrix}, \quad E_{{\mathrm{inc}},2}(r) = \begin{pmatrix} 0 \cr \exp\{-\mathrm{i} \kappa r_1\} \end{pmatrix},
\end{eqnarray}
where $r = (r_1, r_2)^\mathrm{T}$ is the spatial position. The inverse problem of QTAT was studied using one ($E_{{\mathrm{inc}},1}$) and two ($E_{{\mathrm{inc}},1}, E_{{\mathrm{inc}},2}$) electromagnetic excitations. Furthermore, excitation frequencies of $\tilde{f} = 0.3, 1,$ and $3$ GHz were considered. The electric fields were simulated using the Maxwell's equation \eqref{eq:double_curl_impedance_boundary} and a model for the absorbed energy density \eqref{eq:H_instant} using the FE-approximation with edge elements described in Section \ref{subsec:finite_element_approximation}. The FE-approximation was implemented in MATLAB utilising the FEM matrix assembly package published in \cite{anjam2015}. The simulation geometry and the incident electric fields are illustrated in Figure \ref{fig:simulation_illustration}. 

\subsection{Objective function surfaces}
\label{subsec:error_maps}

\begin{figure}[!tbp]
	\centering
	\includegraphics[width=1\textwidth]{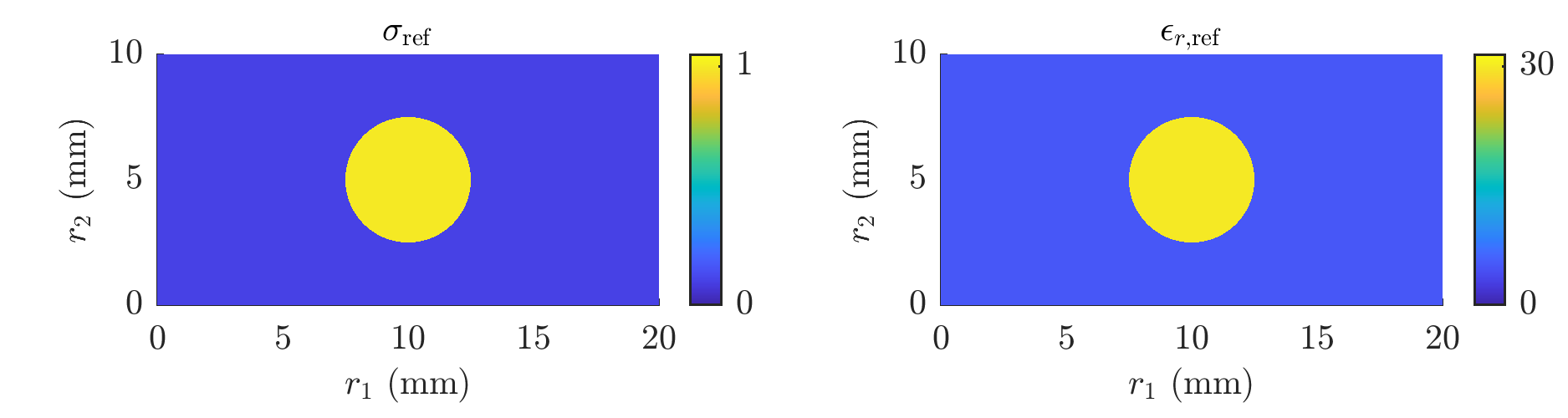}
	\caption{Conductivity $\sigma_{\mathrm{ref}}$ (Sm$^{-1}$) (left image) and relative permittivity $\epsilon_{r,{\mathrm{ref}}}$ (right image) phantoms used in calculating objective function surfaces.}
	\label{fig:error_map_phantom}
\end{figure}

To study the characteristics of the inverse problem, a phantom with one circular inclusion with a radius of $5 \, {\mathrm{mm}}$ illustrated in Figure \ref{fig:error_map_phantom} was considered.
Values of dielectric parameters within the inclusion were varied, and objective function surfaces were calculated as 
\begin{eqnarray}
	\Psi(\sigma_{\mathrm{p}}, \epsilon_{r,{\mathrm{p}}}) = \Vert f(\sigma_{\mathrm{ref}}, \epsilon_{r,{ \mathrm{ref}}}) - f(\sigma_{\mathrm{p}}, \epsilon_{r,{\mathrm{p}}}) \Vert^2,
	\label{eq:objective_function}
\end{eqnarray}
where $f(\sigma_{\mathrm{ref}}, \epsilon_{r,{\mathrm{ref}}})$ is the data computed using the reference parameters and $f(\sigma_{\mathrm{p}}, \epsilon_{r,{\mathrm{p}}})$ is the data computed using permuted inclusion parameters. The background parameters of the phantom were  constants in all simulations, $\sigma_{\mathrm{b}} = 0.1 \, {\mathrm{Sm}}^{-1}$ and $\epsilon_{r,{\mathrm{b}}} = 5$. The inclusion parameters for the reference data were  $\sigma_{\mathrm{ref}} = 1 \, {\mathrm{Sm}}^{-1}$ and $\epsilon_{r,{\mathrm{ref}}} = 30$. For the permuted inclusion parameters, conductivity and relative permittivity values were varied in the ranges $\sigma_{\mathrm{p}} = [0,2] \, {\mathrm{Sm}}^{-1}$ and $\epsilon_{r,{\mathrm{p}}} = [1,70]$, respectively. Data for the objective function surfaces was simulated in a triangular mesh detailed in Table \ref{tab:discretisations}.
\begin{table}[tbp!]
	\caption{\label{tab:discretisations} Number of nodes $N$, elements $J$, and edges $I$ of the FE-meshes used in simulation of the objective function surfaces, simulation of the thermoacoustic data, and solving the inverse problem.}
	\begin{center}
		\begin{tabular}{@{}lrrr} \hline
			& $N$ & $J$ & $I$ \\ \hline
			Objective function surfaces  & 172 373 & 343 244 & 515 616 \\
			Data simulation & 183 353 & 365 204 & 548 556 \\
			Inverse problem     & 11 364 & 22 324   & 33 687 \\ \hline
		\end{tabular}
	\end{center}
\end{table}

\begin{figure}[tbp!]
	\centering
	\includegraphics[width=1\textwidth]{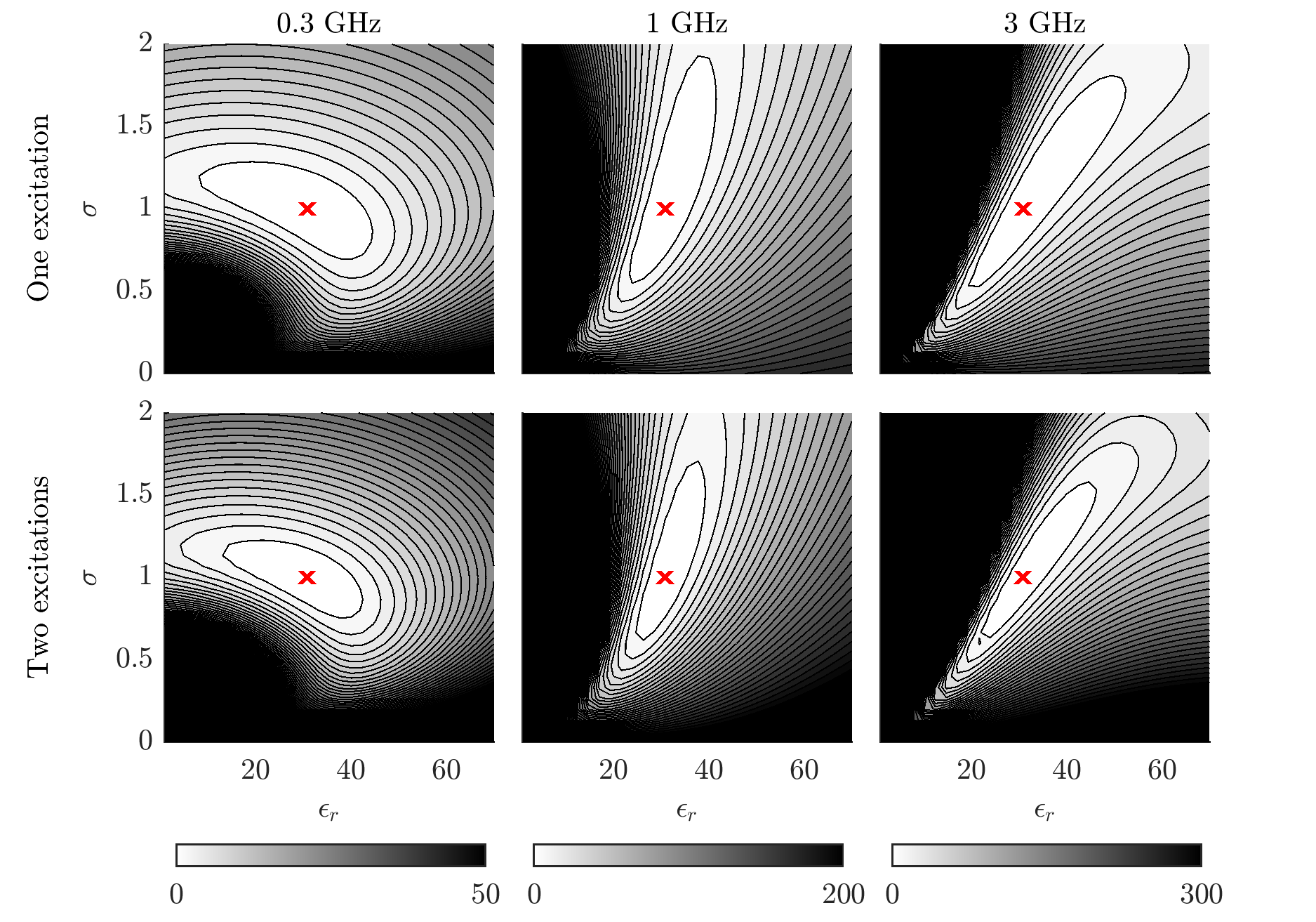}
	\caption{Objective function surfaces calculated using varied conductivity $\sigma$ (Sm$^{-1}$) and relative permittivity $\epsilon_r$ values for one electric field $E_{\mathrm{inc},1}$ (first row) and two electric field $E_{\mathrm{inc},1}$ and $E_{\mathrm{inc},2}$ (second row) excitations at frequencies $\tilde{f} = 0.3, \, 1 \, {\mathrm{and}} \, 3 \, {\mathrm{GHz}}$  (columns from left to right). The reference (true) values of the inclusion are indicated with a red cross. For visualisation, the values of the objective function surfaces are thresholded to maximum values of 50, 200, and 300 for the three frequencies, respectively.}
	\label{fig:error_map_test}
\end{figure}

The surfaces of the objective function \eqref{eq:objective_function} calculated using one electric field excitation $E_{\mathrm{inc},1}$ and two electric field excitations $E_{\mathrm{inc},1}$ and $E_{\mathrm{inc},2}$ at frequencies $\tilde{f} = 0.3,\, 1, \, \mathrm{and}\, 3 \, {\mathrm{GHz}}$ are shown in Figure \ref{fig:error_map_test}. 
As it can be seen, the objective function surfaces differ significantly  on different frequencies. In the case of $\tilde{f} = 0.3 \, {\mathrm{GHz}}$, the objective function surfaces have a relatively wide kidney-shaped minimum. With $\tilde{f} = 1 \, {\mathrm{GHz}}$ and $\tilde{f} = 3 \, {\mathrm{GHz}}$, the minimum is a long narrow region, indicating the possibility of obtaining similar data within different combinations of the dielectric parameters. When comparing the objective function surfaces obtained using one or two electric field excitations, it can be seen that the minimum is more clearly localised when data from two excitations are used. This can indicate that the inverse problem of QTAT is more ill-posed if only one electromagnetic excitation is used. This problem has been studied in \cite{bal2014,ren2023recovering} in a more complex situation than the two-parameter estimation problem considered here.

\subsection{Estimating the electrical conductivity and relative permittivity}
\label{subsubsec:inverse_problem}

\begin{figure}[!tbp]
	\centering
	\includegraphics[width=1\textwidth]{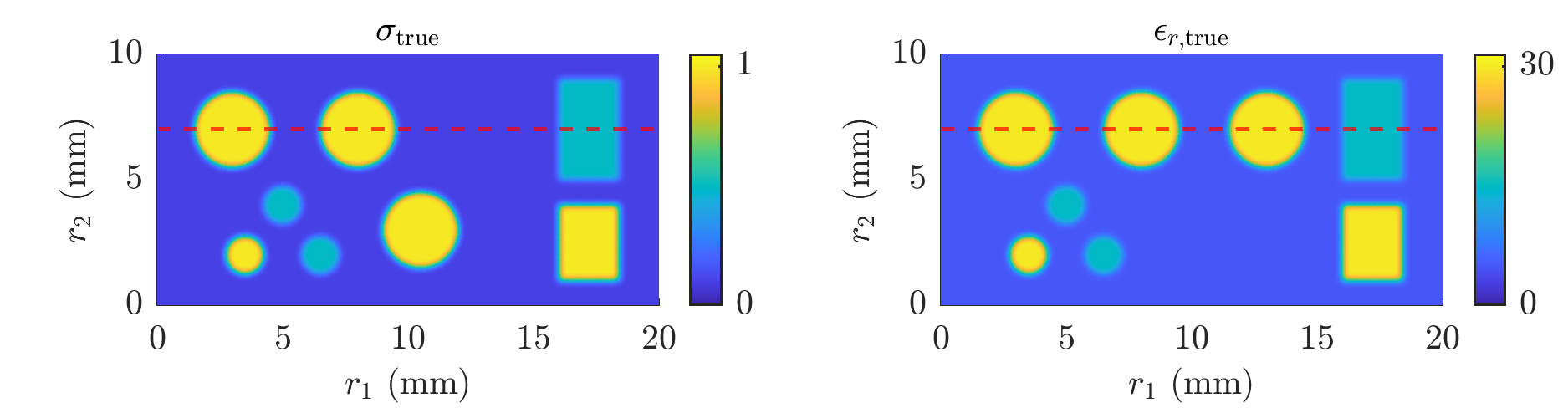}
	\caption{Simulated conductivity $\sigma_{\mathrm{true}}$ (Sm$^{-1}$) (left image) and relative permittivity $\epsilon_{r,{\mathrm{true}}}$ (right image) phantoms used in studying the inverse problem. Location of the line plots used in visualising the results is indicated with a red dashed line.}
	\label{fig:inverse_problem_phantom}
\end{figure}

The inverse problem of QTAT was studied with a numerical phantom consisting of multiple inclusions with different values of electrical conductivity and relative permittivity. The values of the dielectric parameters were chosen to mimic soft biological tissues, such as breast tissue \cite{cheng2018}, with maximum and minimum values set at $\sigma_{\mathrm{min}} = 0.1 \, {\mathrm{Sm}}^{-1}$, $\sigma_{\mathrm{max}} = 1 \, {\mathrm{Sm}}^{-1}$, $\epsilon_{r,\mathrm{min}} = 5$, and $\epsilon_{r,\mathrm{max}} = 30$. The simulated conductivity and relative permittivity distributions are shown in Figure \ref{fig:inverse_problem_phantom}. 

\begin{figure}[!tbp]
	\centering
	\includegraphics[width=1\textwidth]{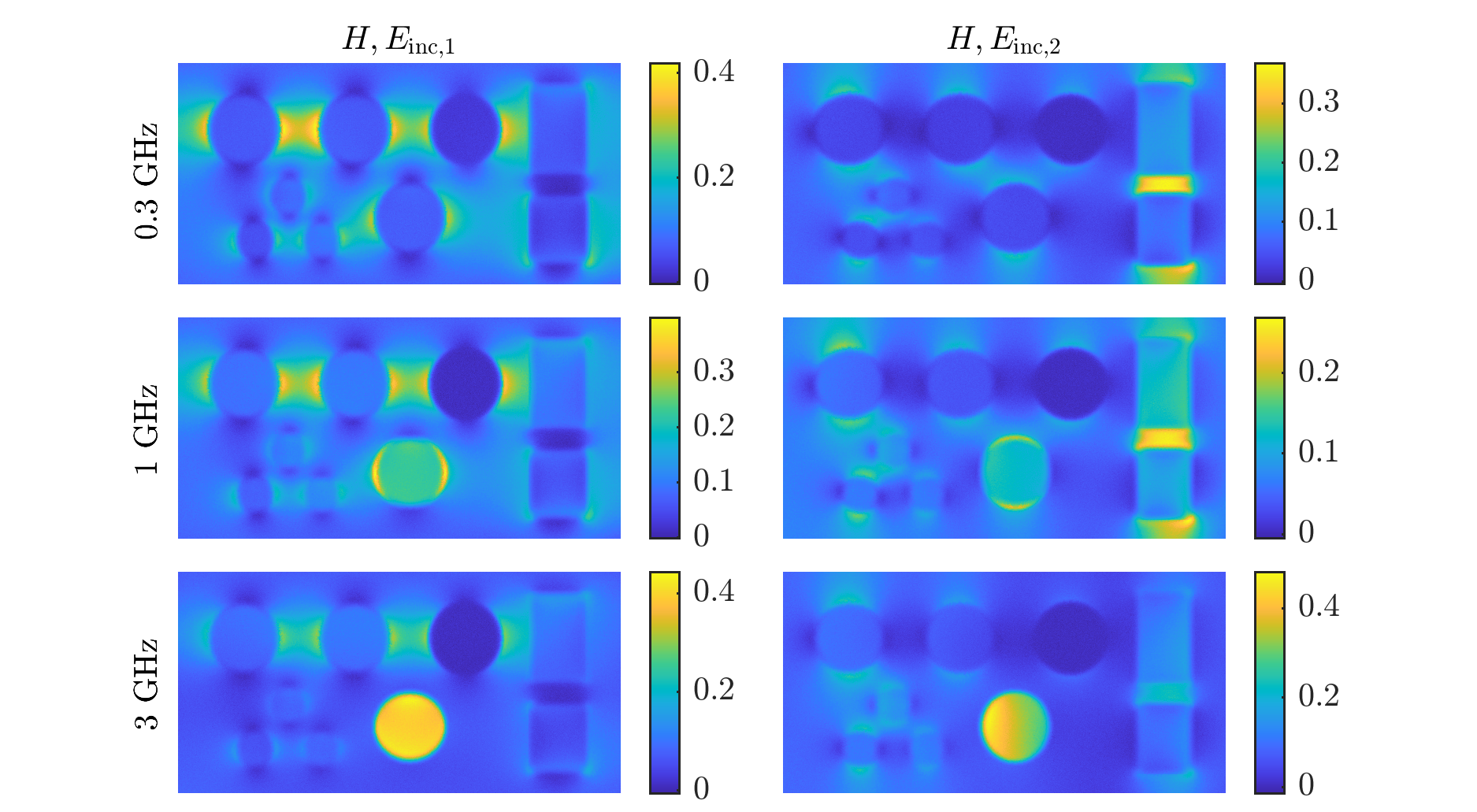}
	\caption{Data $H$ simulated using incident electric fields $E_{\mathrm{inc},1}$ (first column) and $E_{\mathrm{inc},2}$ (second column) at frequencies $\tilde{f}$ = 0.3, 1, and 3 GHz. (rows 1-3).}
	\label{fig:data_plot}
\end{figure}

Absorbed energy density data was simulated in a triangular simulation mesh detailed in Table \ref{tab:discretisations} using both one and two electromagnetic excitations at three different excitation frequencies. After data simulation, Gaussian distributed zero-mean noise with a standard deviation set to $1 \, \% $ of the maximum simulated amplitude was added to the data. Noisy data in the simulation mesh for the incident electric fields $E_{\mathrm{inc},1}$ and $E_{\mathrm{inc},2}$ at frequencies $\tilde{f} = 0.3,\, 1, \, \mathrm{and}\, 3 \, {\mathrm{GHz}}$ are shown in Figure \ref{fig:data_plot}. 
For the inverse problem, the data was linearly interpolated to a coarser reconstruction mesh described in Table \ref{tab:discretisations}.

In the inverse problem, the conductivity $\sigma$ and relative permittivity $\epsilon_r$ were estimated simultaneously by computing the MAP estimates \eqref{eq:map_optimisation} from simulated data. The noise was modelled as uncorrelated  Gaussian distributed noise with a zero mean and standard deviation of $1 \, \%$  of the maximum amplitude of the (noisy) simulated data. In this work, Ornstein-Uhlenbeck prior model \cite{rasmussen2006} was used for the unknown parameters, where the covariance is defined as 
\begin{eqnarray}
	\Gamma_{x} =\begin{bmatrix} \tilde{\sigma}^2_\sigma \Pi & 0 \cr 0 & \tilde{\sigma}^2_{\epsilon_r} \Pi \end{bmatrix},
\end{eqnarray}
where $\tilde{\sigma}_\sigma$ is the standard deviation of the conductivity and $\tilde{\sigma}_{\epsilon_r}$ is the standard deviation of the relative permittivity. Furthermore, $\Pi$ is defined as
\begin{eqnarray}
	\Pi_{j_1, j_2} = \exp \left\{ -\frac{\Vert r_{j_1} - r_{j_2} \Vert}{\ell} \right\},
\end{eqnarray}
where $r_{j_1}$ and $r_{j_2}$ denote the locations of the triangular elements, and $\ell$ is the characteristic length scale controlling the spatial correlation. The expected values of the prior were set as $\eta_\sigma = \frac{1}{2}(\sigma_{\mathrm{max}} + \sigma_{\mathrm{min}})$ and $\eta_{\epsilon_r} = \frac{1}{2}(\epsilon_{r,{\mathrm{max}}} + \epsilon_{r,{\mathrm{min}}})$, and standard deviations were set as $\tilde{\sigma}_\sigma = \frac{1}{2}(\sigma_{\mathrm{max}} - \sigma_{\mathrm{min}})$ and $\tilde{\sigma}_{\epsilon_r} = \frac{1}{2}(\epsilon_{r,{\mathrm{max}}} - \epsilon_{r,{\mathrm{min}}})$ where $\sigma_{\mathrm{min}}$, $\sigma_{\mathrm{max}}$, $\epsilon_{r,{\mathrm{min}}}$ and $\epsilon_{r,{\mathrm{max}}}$ are the minimum and maximum values of the simulated conductivity and relative permittivity phantom. The characteristic length scale was set as $\ell = 1  \, {\mathrm{mm}}$. 

The minimisation problem was solved using the Gauss-Newton method \eqref{eq:minimisation_iteration}--\eqref{eq:GN_search_direction} equipped with constraints  $\sigma \geq 0$ and $\epsilon_r \geq 1$, and a line-search algorithm for the selection of the step length $\alpha$. The initial guess of the Gauss-Newton algorithm was set at the background values $\sigma_{\mathrm{min}}$ and $\epsilon_{r,{\mathrm{min}}}$ and the minimisation problem was computed for 15 iterations. The standard deviation and credibility intervals were computed using \eqref{eq:covariance} and \eqref{eq:credibility_interval}. 

\begin{figure}[!tbp]
	\centering
	\includegraphics[width=1\textwidth]{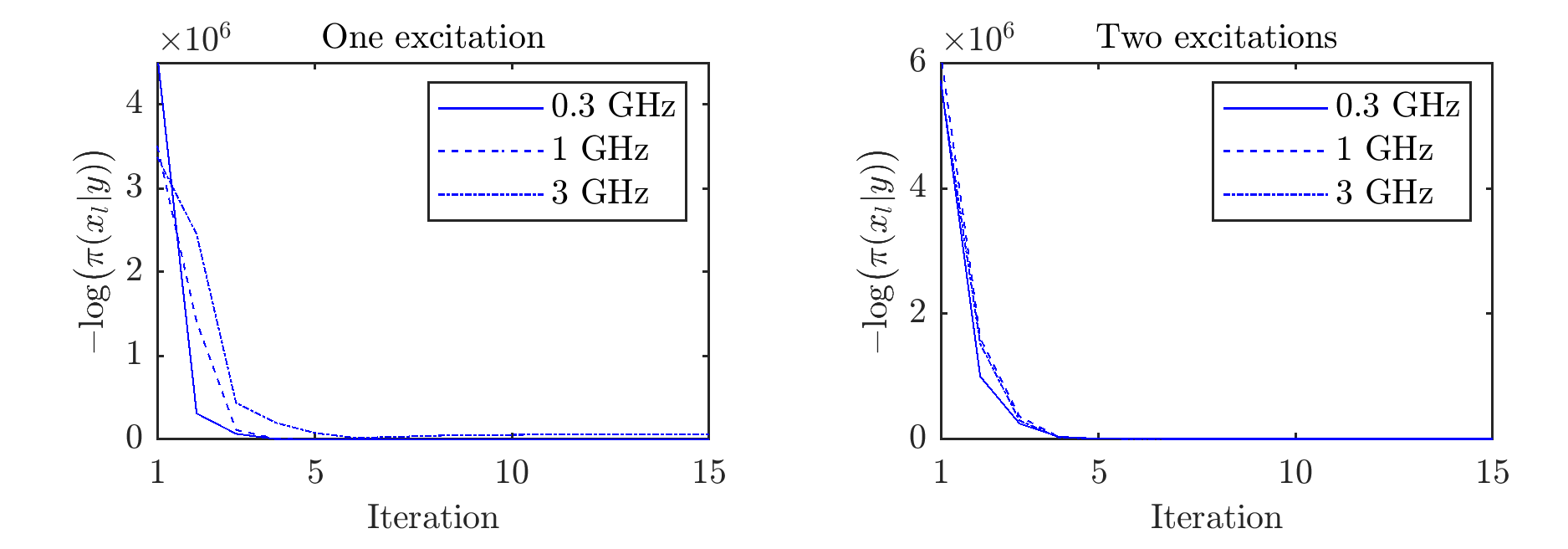}
	\caption{Objective function values of the MAP estimation problem \eqref{eq:map_optimisation} using one electric field excitation $E_{\mathrm{inc},1}$  (left image) and two  electric field excitations   $E_{\mathrm{inc},1}$ and $E_{\mathrm{inc},2}$ (right image) at frequencies $\tilde{f} = 0.3, 1,$ and $3$ GHz.}
	\label{fig:residual_plot}
\end{figure}

\begin{figure}[tbp!]
	\centering
	\includegraphics[width=1\textwidth]{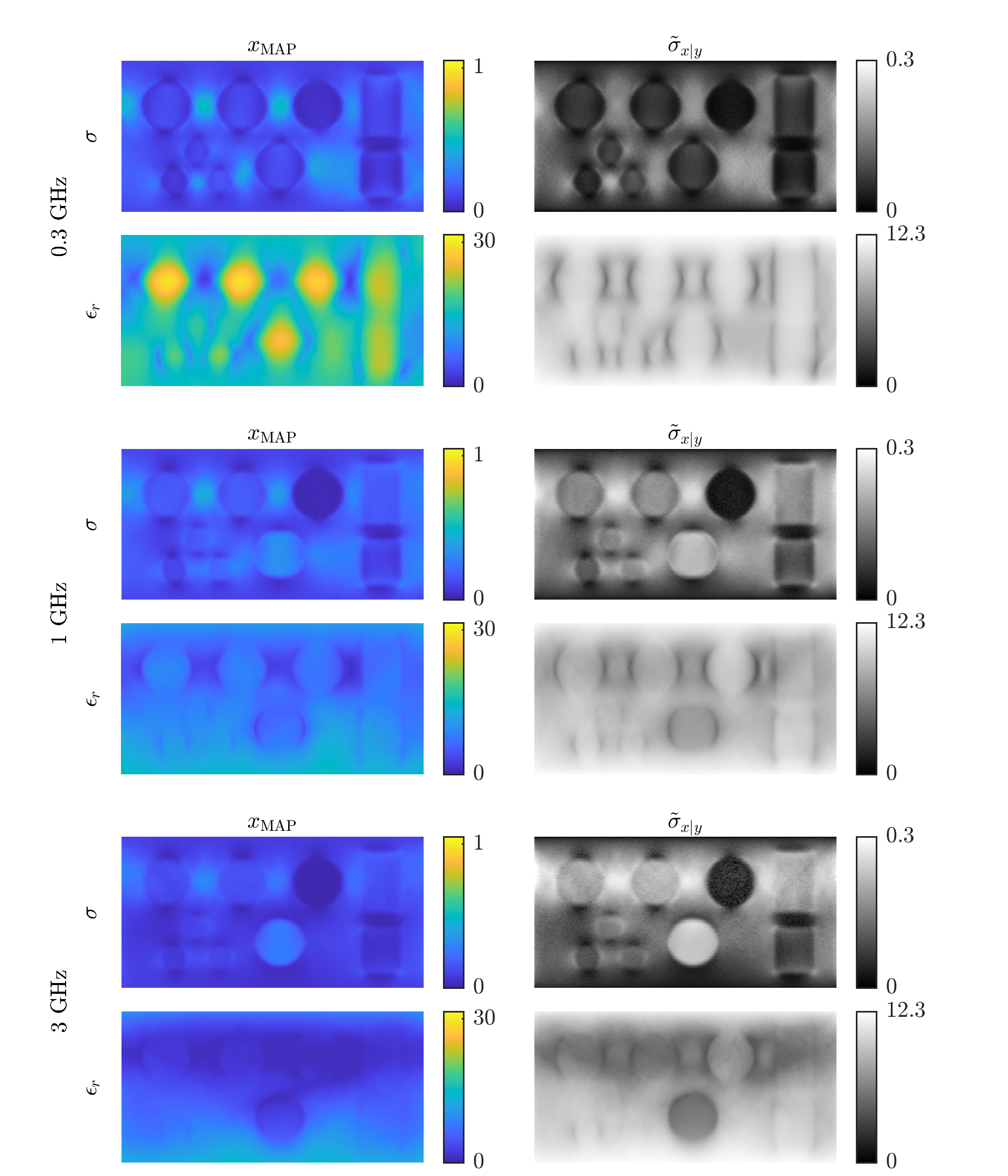}
	\caption{MAP-estimates $x_{\mathrm{MAP}}$ (first column) and standard deviations $\tilde{\sigma}_{x\vert y}$ (second column) for conductivity $\sigma$ (Sm$^{-1}$) and relative permittivity $\epsilon_r$ obtained using one  electric field  excitation $E_{\mathrm{inc},1}$ at frequencies 0.3 GHz (rows 1-2), 1 GHz (rows 3-4), and 3 GHz (rows 5-6).}
	\label{fig:recos_oneill}
\end{figure}

\begin{figure}[tbp!]
	\centering
	\includegraphics[width=1\textwidth]{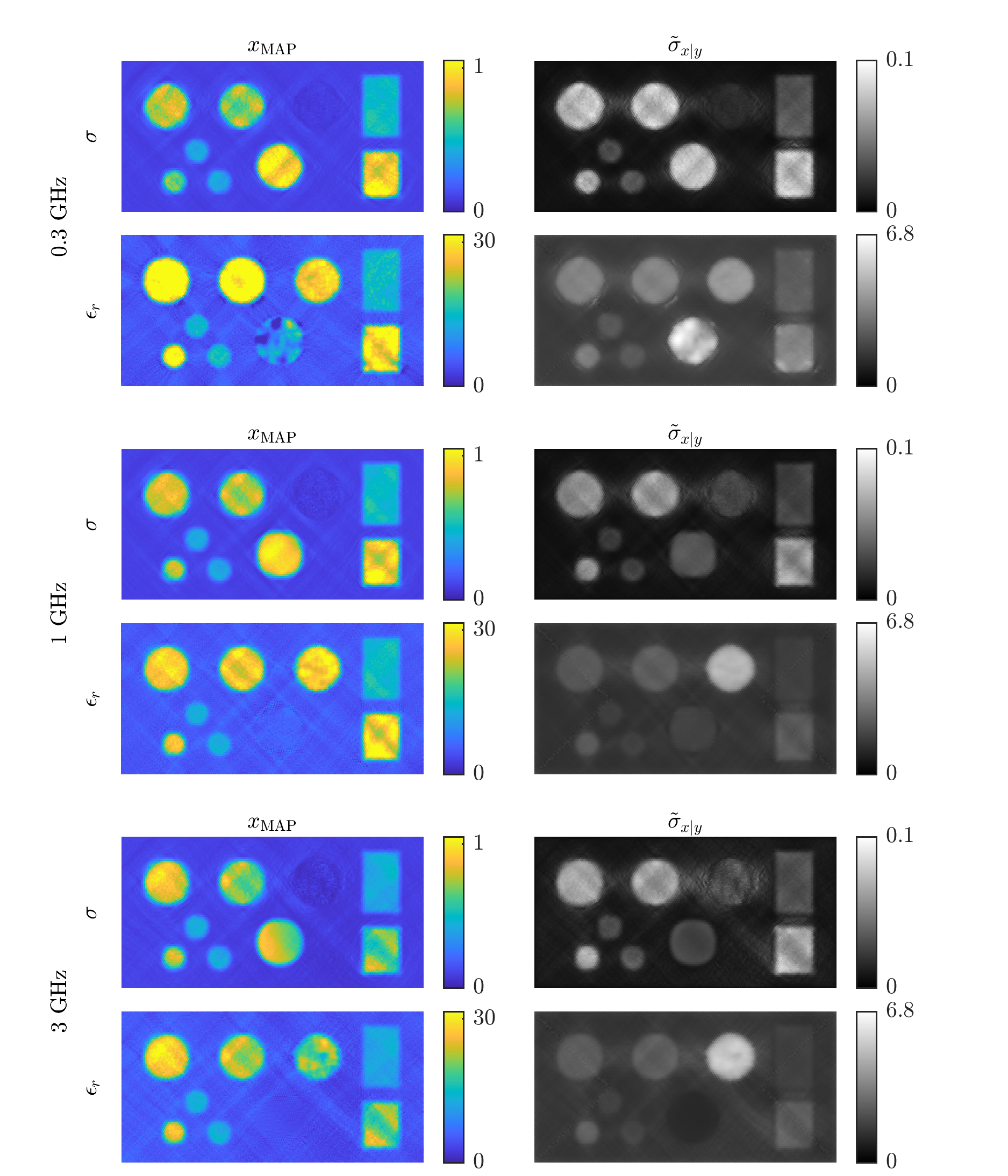}
	\caption{MAP-estimates $x_{\mathrm{MAP}}$ (first column) and standard deviations $\tilde{\sigma}_{x\vert y}$ (second column) for conductivity $\sigma$ (Sm$^{-1}$) and relative permittivity $\epsilon_r$ obtained using two  electric field  excitations $E_{\mathrm{inc},1}$ and $E_{\mathrm{inc},2}$  at frequencies 0.3 GHz (rows 1-2), 1 GHz (rows 3-4), and 3 GHz (rows 5-6).}
	\label{fig:recos_twoill}
\end{figure}

\begin{figure}[tbp!]
	\centering
	\includegraphics[width=\textwidth]{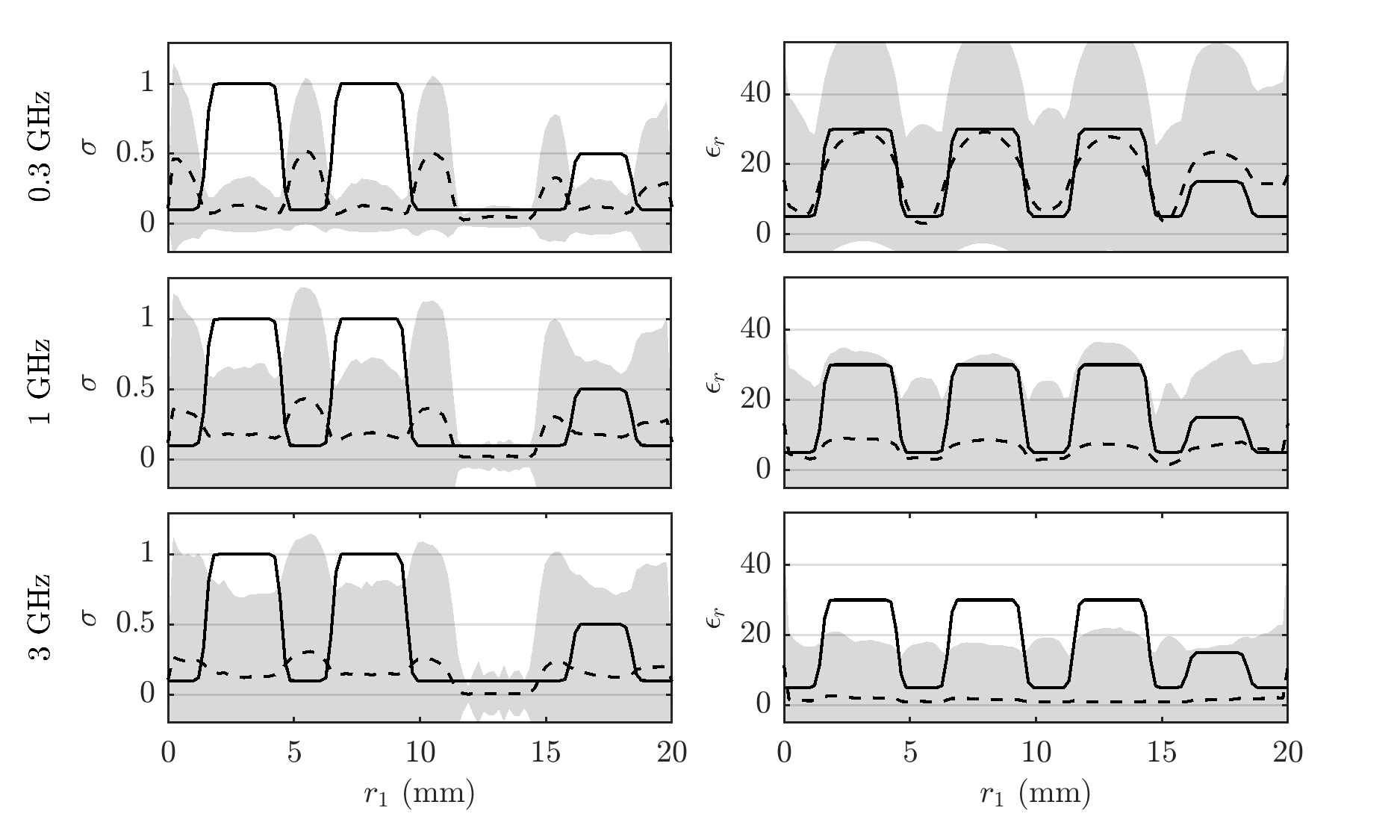}
	\caption{Estimated conductivity $\sigma$ (Sm$^{-1}$) (first column, dashed line) and relative permittivity $\epsilon_r$ (second column, dotted line) and $\pm$3 standard deviation credibility intervals (grey filled area) obtained using one  electric field  excitation $E_{\mathrm{inc},1}$. Results for frequencies  0.3 GHz, 1 GHz, and 3 GHz are shown on rows 1-3. The true values are shown using a black solid line.  }
	\label{fig:cred_oneill}
\end{figure}

\begin{figure}[tbp!]
	\centering
	\includegraphics[width=\textwidth]{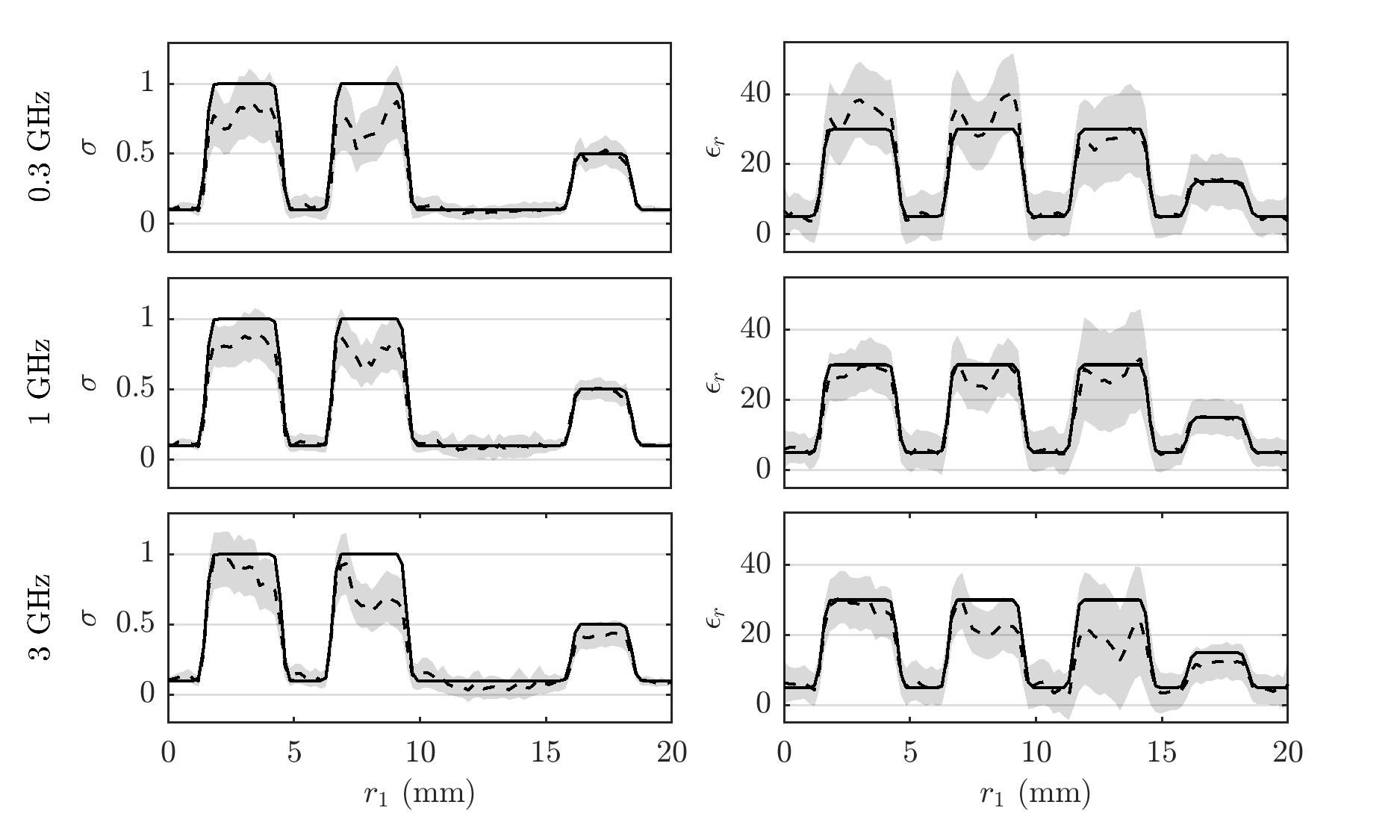}
	\caption{Estimated conductivity $\sigma$ (Sm$^{-1}$) (first column, dashed line) and relative permittivity $\epsilon_r$ (second column, dotted line) and $\pm$3 standard deviation credibility intervals (grey filled area) obtained using two  electric field  excitations $E_{\mathrm{inc},1}$ and $E_{\mathrm{inc},2}$. Results for frequencies  0.3 GHz, 1 GHz, and 3 GHz are shown on rows 1-3. The true values are shown using a black solid line.  }
	\label{fig:cred_twoill}
\end{figure}

The objective function values of the minimisation problem \eqref{eq:map_optimisation} using one electric field excitation $E_{\mathrm{inc},1}$ and two electric field excitations $E_{\mathrm{inc},1}$ and $E_{\mathrm{inc},2}$ at frequencies $\tilde{f} = 0.3,\, 1, \, \mathrm{and}\, 3 \, {\mathrm{GHz}}$ are shown in Figure \ref{fig:residual_plot}. As it can be observed, the optimisation problem converged in all cases. 
The MAP estimates and standard deviations obtained using one electric field excitation $E_{\mathrm{inc},1}$ and two electric field excitations $E_{\mathrm{inc},1}$ and $E_{\mathrm{inc},2}$ at frequencies $\tilde{f} = 0.3,\, 1, \, \mathrm{and}\, 3 \, {\mathrm{GHz}}$ are shown in Figures \ref{fig:recos_oneill} and \ref{fig:recos_twoill}. The corresponding line plots and credibility intervals are shown in Figures \ref{fig:cred_oneill} and \ref{fig:cred_twoill}. 

As it can be seen from Figures \ref{fig:recos_oneill} and \ref{fig:cred_oneill}, the estimates for conductivity and relative permittivity in the case of one electromagnetic excitation are inaccurate at all frequencies. Furthermore, the true values do not lie within the credibility intervals. 

The results obtained using two electromagnetic excitations (Figures \ref{fig:recos_twoill} and \ref{fig:cred_twoill}) are significantly more accurate compared to the case of one excitation. The inclusions are now clearly visible in both reconstructed conductivity and relative permittivity images and their values are close to the true values. Some cross talk can, however, be observed between the conductivity and relative permittivity images in the inclusion locations where only either conductivity or relative permittivity is changed from the background values. Furthermore, some slight streak-like artefacts can be seen in the MAP estimates at all frequencies. Looking at the standard deviations and credibility intervals, it can be observed that the standard deviation generally increases with increasing conductivity and relative permittivity values. However, the true conductivity and relative permittivity values do not lie within the $\pm$3 standard deviation credibility interval everywhere, indicating that the credible intervals cannot be considered reliable in those regions.

\section{Discussions}
\label{sec:discussion}

The results show that using only one electric field excitation results in inaccurate estimates of the dielectric parameters. This is most likely due to the non-uniqueness of the problem when only one excitation is used. Uniqueness of the QTAT problem has been previously studied in \cite{bal2014,ren2023recovering}, and our numerical simulations support these findings. On the other hand, when two electric field excitations are used, the MAP estimates resemble the true parameter values, with only small artefacts and cross-talk between the estimates. The true values do not, however, lie within the credible intervals everywhere in the domain, and thus the methodology could require, for example, improvement of the measurement geometry, modelling of errors in the data likelihood, or the use of more precise prior information. 

In this work, the incident electric fields were modelled as linearly polarised perpendicularly oriented plane waves. Polarisation and orientation of the electromagnetic waves can, however, have a significant effect on the on the absorbed energy density data, and thus effect the solution of the inverse problem. The effect of polarisation on thermoacoustic data has been studied, for example, by the authors in \cite{he2017}, where significant differences in the absorbed energy density were observed between linearly and circularly polarised excitations. We further believe that the streak-like artefacts in the MAP estimates, that seem to be oriented in a 45 degree angle with respect to the excitations, can result from the use of two perpendicular excitations. Therefore, to improve the applicability of the methodology, the effect of different factors related to the incident electric field excitations, such as polarisation, and the number and angle of the incident fields, should studied more extensively.

In this work, the dielectric parameters of the phantoms for the inverse problem were held constant for all studied frequencies to allow for straightforward comparison. In practice, however, the dielectric parameters are dependent of the frequency of the electromagnetic excitation, which should be taken into account in future work. In addition, the imaged target was modelled in an infinitely sized medium and the excitation was assumed as perpendicular plane waves. These assumption might not be accurate in realistic measurement setups that are typically composed of a metallic casing and an electromagnetic source such as a waveguide or an antenna. To move towards a more realistic framework, modelling of the full measurement setup including accurate modelling of the source should be considered. Furthermore, the experimental thermoacoustic measurement systems commonly employ electromagnetic sources with pulse lengths from tens of nanoseconds to few microsecond \cite{cui2017review}. The use of the forward model \eqref{eq:H_instant} can, therefore, lead to modelling errors when applied to data from experimental systems. Therefore, this work should be extended by the temporal variation of the excitation and full thermal effects described in \cite{akhoyari2017}.

\section{Conclusions}
\label{sec:conclusions}

In this work, an approach for simultaneous estimation of electrical conductivity and relative permittivity in QTAT was studied using numerical simulations. MAP estimates were computed from absorbed energy density data, and the reliability of the estimates was evaluated using Laplace's approximation. The problem was studied using one and two electric field excitations at different frequencies. The results show that the electrical conductivity and permittivity can be simultaneously estimated in QTAT. However, the problem can suffer non-uniqueness, that could be overcome using multiple electromagnetic excitations.

\section*{Acknowledgements}
This work was supported by the European Research Council (ERC) under the European Union’s Horizon 2020 Research and Innovation Program (Grant Agreement No. 101001417-QUANTOM) and the Research Council of Finland (Centre of Excellence in Inverse Modelling and Imaging grant 353086, and Flagship of Advanced Mathematics for Sensing Imaging and Modelling grant 358944, and project grant 321761). The authors wish to acknowledge CSC – IT Center for Science, Finland, for computational resources. The authors would like to thank Amelie Litman for useful discussions.

\bibliographystyle{plain}
\bibliography{biblio}

\begin{thebibliography}{10}

\bibitem{akhoyari2017}
H.~Akhouayri, M.~Bergounioux, A.~Da Silva, P.~Elbau, A.~Litman, and
  L.~Mindrinos.
\newblock Quantitative thermoacoustic tomography with microwaves sources.
\newblock {\em Journal of Inverse and Ill-posed Problems}, 25(6):703--717,
  2017.

\bibitem{ammari2013}
H.~Ammari, J.~Garnier, W.~Jiang, and L.~H. Nguyen.
\newblock Quantitative thermo-acoustic imaging: An exact reconstruction
  formula.
\newblock {\em Journal of Differential Equations}, 254(3):1375--1395, 2013.

\bibitem{anjam2015}
I.~Anjam and J.~Valdman.
\newblock Fast matlab assembly of fem matrices in 2d and 3d: Edge elements.
\newblock {\em Applied Mathematics and Computation}, 267:252--263, 2015.

\bibitem{bal2011}
G.~Bal, K.~Ren, G.~Uhlmann, and T.~Zhou.
\newblock Quantitative thermo-acoustics and related problems.
\newblock {\em Inverse Problems}, 27:055007, 2011.

\bibitem{bal2014}
G.~Bal and T.~Zhou.
\newblock Hybrid inverse problems for a system of maxwell’s equations.
\newblock {\em Inverse Problems}, 30:055013, 2014.

\bibitem{thesis_bonazolli}
M.~Bonazolli.
\newblock {\em Efficient high order and domain decomposition methods for the
  time-harmonic Maxwell's equations}.
\newblock PhD thesis, Universit\'e C\^ote D'Azur, 2017.

\bibitem{book_calvetti}
D.~Calvetti and E.~Somersalo.
\newblock {\em Bayesian Scientific Computing}.
\newblock Springer-Verlag, 2024.

\bibitem{cheng2018}
Y.~Cheng and M.~Fu.
\newblock Dielectric properties for non‐invasive detection of normal, benign,
  and malignant breast tissues using microwave theories.
\newblock {\em Thorasic Cancer}, 9(4):459--465, 2018.

\bibitem{cox2012}
B.~Cox, J.~G. Laufer, S.~R. Arridge, and P.~C. Beard.
\newblock Quantitative spectroscopic photoacoustic imaging:a review.
\newblock {\em Journal of Biomedical Optics}, 17(6):061202, 2012.

\bibitem{cristofol2023carleman}
M.~Cristofol, S.~Li, and Y.~Shang.
\newblock Carleman estimates and some inverse problems for the coupled
  quantitative thermoacoustic equations by partial boundary layer data. {P}art
  {II}: Some inverse problems.
\newblock {\em Mathematical Methods in the Applied Sciences},
  46(12):13304--13319, 2023.

\bibitem{cui2017review}
Y.~Cui, C.~Yuan, and Z.~Ji.
\newblock A review of microwave-induced thermoacoustic imaging: Excitation
  source, data acquisition system and biomedical applications.
\newblock {\em Journal of Innovative Optical Health Sciences}, 10(04):1730007,
  2017.

\bibitem{fallon2009rf}
D.~Fallon, L.~Yan, G.W. Hanson, and S.K. Patch.
\newblock Rf testbed for thermoacoustic tomography.
\newblock {\em Review of Scientific Instruments}, 80(6), 2009.

\bibitem{book_harrington}
R.~F. Harrington.
\newblock {\em Time-Harmonic Electromagnetic Fields}.
\newblock IEEE Press, USA, 2001.

\bibitem{he2017}
Y.~He, C.~Liu, and L.~Lin L.~V. Wang.
\newblock Comparative effects of linearly and circularly polarized illumination
  on microwave-induced thermoacoustic tomography.
\newblock {\em IEEE Antennas and Wireless Propagation Letters}, 16:1593--1596,
  2017.

\bibitem{al2020direct}
H.~Al Jebawy and A.~El Badia.
\newblock Direct algorithm for reconstructing small absorbers in thermoacoustic
  tomography problem from a single data.
\newblock {\em Inverse Problems}, 36(6):065010, 2020.

\bibitem{book_kaipio}
J.~Kaipio and E.~Somersalo.
\newblock {\em Statistical and Computational Inverse problems}.
\newblock Springer-Verlag, 2006.

\bibitem{kruger1999thermoacoustic}
R.A. Kruger, K.K. Kopecky, A.M. Aisen, D.R. Reinecke, G.A. Kruger, and
  W.L.~Kiser Jr.
\newblock Thermoacoustic {CT} with radio waves: A medical imaging paradigm.
\newblock {\em Radiology}, 211(1):275--278, 1999.

\bibitem{kuchment2011bookchapter}
P.~Kuchment and L.~Kunyansky.
\newblock {\em Mathematics of photoacoustic and thermoacoustic tomography},
  chapter~19, pages 817--865.
\newblock Springer New York, 2011.

\bibitem{li2008}
C.~Li, M.~Pramanik, G.~Ku, and L.~V. Wang.
\newblock Image distortion in thermoacoustic tomography caused by microwave
  diffraction.
\newblock {\em Physical Review E}, 77:031923, 2008.

\bibitem{li2009}
C.~Li and L.~V. Wang.
\newblock Photoacoustic tomography and sensing in biomedicine.
\newblock {\em Physics in Medicine \& Biology}, 54:R59--R97, 2009.

\bibitem{book_li}
J.~Li and Y.~Huang.
\newblock {\em Time-Domain Finite Element Methods for Maxwell's Equations in
  Metamaterials}.
\newblock Springer-Verlag, Germany, 2013.

\bibitem{book_monk}
P.~Monk.
\newblock {\em Finite Element Methods for Maxwell's Equations}.
\newblock Claredon Press, United Kingdom, 2003.

\bibitem{nocedal_book}
J.~Nocedal and S.~J. Wright.
\newblock {\em Numerical Optimization}.
\newblock Springer, the United States of America, 2006.
\newblock Second Edition.

\bibitem{ogunlade2012quantitative}
O.~Ogunlade, B.~Cox, and P.~Beard.
\newblock Quantitative thermoacoustic image reconstruction of conductivity
  profiles.
\newblock In {\em Photons Plus Ultrasound: Imaging and Sensing 2012}, volume
  8223, pages 142--151. SPIE, 2012.

\bibitem{poudel2019survey}
J.~Poudel, Y.~Lou, and M.A. Anastasio.
\newblock A survey of computational frameworks for solving the acoustic inverse
  problem in three-dimensional photoacoustic computed tomography.
\newblock {\em Physics in Medicine \& Biology}, 64(14):14TR01, 2019.

\bibitem{pulkkinen2016}
A.~Pulkkinen, B.~T. Cox, S.~R. Arridge, H.~Goh, J.~P. Kaipio, and T.~Tarvainen.
\newblock Direct estimation of optical parameters from photoacoustic time
  series in quantitative photoacoustic tomography.
\newblock {\em IEEE Transactions on Medical Imaging}, 35(11):2497--2508, 2016.

\bibitem{rasmussen2006}
C.~E. Rasmussen and C.~K.~I. Williams.
\newblock {\em Gaussian Processes for Machine Learning}.
\newblock MIT Press, 2006.

\bibitem{razansky2010near}
D.~Razansky, S.~Kellnberger, and V.~Ntziachristos.
\newblock Near-field radiofrequency thermoacoustic tomography with mpulse
  excitation.
\newblock {\em Medical Physics}, 37(9):4602--4607, 2010.

\bibitem{ren2023recovering}
K.~Ren and N.~Soedjak.
\newblock Recovering coefficients in a system of semilinear helmholtz equations
  from internal data.
\newblock {\em arXiv preprint}, arXiv:2307.01385v1 [math.AP], 2023.

\bibitem{book_tarantola}
A.~Tarantola.
\newblock {\em Inverse Problem Theory and Methods for Model Parameter
  Estimation}.
\newblock Society for Industrial and Applied Mathematics, 2005.

\bibitem{tarvainen2024quantitative}
T.~Tarvainen and B.~Cox.
\newblock Quantitative photoacoustic tomography: modeling and inverse problems.
\newblock {\em Journal of Biomedical Optics}, 29(S1):S11509--S11509, 2024.

\bibitem{wang2011bookchapter}
K.~Wang and M.A. Anastasio.
\newblock {\em Photoacoustic and Thermoacoustic Tomography: Image Formation
  Principles}, chapter~18, pages 781--815.
\newblock Springer New York, 2011.

\bibitem{wang1999microwave}
L.V. Wang, X.~Zhao, H.~Sun, and G.~Ku.
\newblock Microwave-induced acoustic imaging of biological tissues.
\newblock {\em Review of Scientific Instruments}, 70(9):3744--3748, 1999.

\bibitem{thesis_watson}
F.~Watson.
\newblock {\em Better imaging for landmine detection: An exploration of 3D
  full-wave inversion for ground-penetrating radar}.
\newblock PhD thesis, The University of Manchester, 2016.

\end{thebibliography}

\end{document}